# A new smart-cropping pipeline for prostate segmentation using deep learning networks[1]

Dimitrios G. Zaridis, Eugenia Mylona, Nikolaos S. Tachos, Kostas Marias, Nikolaos Papanikolaou, Manolis Tsiknakis and Dimitrios I. Fotiadis, *Fellow Member IEEE*

*Abstract*— Prostate segmentation from magnetic resonance imaging (MRI) is a challenging task. In recent years, several network architectures have been proposed to automate this process and alleviate the burden of manual annotation. Although the performance of these models has achieved promising results, there is still room for improvement before these models can be used safely and effectively in clinical practice. One of the major challenges in prostate MR image segmentation is the presence of class imbalance in the image labels where the background pixels dominate over the prostate. In the present work we propose a DL-based pipeline for cropping the region around the prostate from MRI images to produce a more balanced distribution of the foreground pixels (prostate) and the background pixels and improve segmentation accuracy. The effect of DL-cropping for improving the segmentation performance compared to standard center-cropping is assessed using five popular DL networks for prostate segmentation, namely U-net, U-net+, Res Unet++, Bridge U-net and Dense U-net. The proposed smart-cropping outperformed the standard center cropping in terms of segmentation accuracy for all the evaluated prostate segmentation networks. In terms of Dice score, the highest improvement was achieved for the U-net+ and ResU-net++ architectures corresponding to 8.9% and 8%, respectively.

I. INTRODUCTION

Prostate cancer (PCa) is the second most prevalent type of cancer and the fifth leading cause of cancer death in men [1]. Despite the high prevalence of the disease, with appropriate treatment, the 5-year survival rate is up to 98% [2].

Magnetic Resonance Imaging (MRI) is one of the most reliable prostate imaging modalities and non-invasive diagnostic methods of prostate cancer [3]. Multi-parametric MRI, in particular, is emerging as a clinically useful tool for detecting, localizing and staging prostate cancer. Segmentation of the prostate from MRI is a fundamental step of the medical image analysis for diagnosis, surgery and therapy. Several clinical tasks, including cancer detection, localization and staging, treatment planning, medical intervention and targeted MRI-transrectal ultrasound (TRUS) fusion guided biopsy and therapy are highly dependent on an accurate delineation of the prostate in imaging data [4–6].

The wide range of inter-individual shape variation of the prostate and distortions on MRI images due to field inhomogeneity, make the segmentation of the prostate a challenging task [7]. Current practices include slice-by-slice manual contouring by a specialist, which is a time and labor intensive task as well as susceptible to intra- and inter-observer variability. By the automation of that technique faster treatment planning is possible while the quality is not compromised. To alleviate the burden of manual annotation, several methods have been proposed to automatically segment the prostate gland and other regions of interest (ROI). The classical method of image segmentation is based on edge detection filters and mathematical algorithms. Alternatively, atlas-based registration has been used for organ segmentation [8]. Today, image segmentation consists the main target for Deep Learning (DL) approaches in medical imaging [9].

Deep learning has witnessed a tremendous amount of attention over the last decade enabling computers to discover complicated patterns in large datasets. Motivated by the success of DL in a variety of machine learning applications such as computer vision and language modeling, researchers in the medical image field have also applied DL-based approaches for medical image segmentation proposing a plethora of network architectures. Specifically, supervised classification methods based on convolutional neural networks (CNNs), have pushed forward the field of medical imaging for segmenting the anatomy of interest, with U-net being a major breakthrough [10]. U-shaped networks have produced higher accuracies for automatic prostate segmentations from T2-weighted MRIs, compared with alternative segmentation approaches [11].

In prostate segmentation tasks, the performance of these models has achieved promising results. Nevertheless, further improvement is required to guarantee their safe and effective application in clinical practice [12]. One of the major challenges in prostate MR image segmentation is the presence of class imbalance in the image labels where the background pixels dominate over the prostate [13]. The issue arises mainly from the usage of cross entropy (CE) – the most commonly used loss function, as it is well known in literature that CE has difficulty in handling class imbalance. As a result, training with imbalanced data can cause an unstable segmentation network, which is biased towards the majority class (background pixels). The utilization of resampling in regards

*This work is supported by the ProCancer-I project that has received funding from the European Union's Horizon 2020 research and innovation program under grant agreement No 952159. This work reflects only the author's view. The Commission is not responsible for any use that may be made of the information it contains.

D.G. Zaridis, E. Mylona, N.S. Tachos are with the Dept. of Biomedical Research, FORTH-IMBB, Ioannina, Greece and the Unit of Medical Technology and Intelligent Information Systems, Dept. of Materials Science and Engineering, University of Ioannina, Ioannina, GR 45110, Greece (e-mail: dimzaridis@gmail.com, ntachos@gmail.com).

K. Marias is with the Institute of Computer Science, FORTH, Heraklion, Greece and the Dept. of Electrical and Computer Engineering, Hellenic Mediterranean University, Greece.

N. Papanikolaou is with the Centre for the Unknown, Champalimaud Foundation, Clinical Computational Imaging Group, Lisbon, Portugal.

M. Tsiknakis is with the Institute of Computer Science, Foundation for Research and Technology Hellas (FORTH), Heraklion, Crete, Greece.

D.I. Fotiadis is with the Dept of Biomedical Research, Institute of Molecular Biology and Biotechnology, FORTH, Ioannina, Greece and the Dept. of Materials Science and Engineering, Unit of Medical Technology and Intelligent Information Systems, University of Ioannina, GR 45110, Ioannina, Greece (corresponding author, e-mail: fotiadis@uoi.gr).

of images and the application of the right loss function are essential strategies to overcome the issue [14]. For instance, center-cropping the image prior to network training has been used in order to reduce the size of background pixels with respect to the ROI [15,16]. Usually, a random offset is added to avoid location bias, where a ROI is always expected to be at the center of the image. This may be efficient for stable and large ROIs, such as the thorax [15], but in case of prostate segmentation, where the MRI depicts the whole pelvic anatomy, center-cropping endangers placing the prostate far away from the center, leading to inaccurate segmentation [17].

In the present work we propose a DL-based pipeline to accurately crop prostate MRI images while optimizing a balanced distribution of the foreground pixels (prostate) and the background pixels to improve prostate segmentation accuracy. Our contributions in this work are (a) the development of a smart-cropping technique that allows to accurately crop the prostate gland from multiparametric-MR images and (b) a comprehensive comparison of different DL networks for prostate segmentation to evaluate the effect of smart-cropping for improving their performance.

## II. MATERIALS AND METHODS

### A. Dataset

In the proposed work, we used the Prostate-3T dataset which is publicly available at The Cancer Imaging Archive (TCIA) [18]. and it contains 60 patients, 30 for training with annotations and 30 for testing without annotations. The annotations provided by the abovementioned dataset are masks from the central prostate gland, the prostate's peripheral zone and the seminal vesicles. The acquisition method of the data is the T2 weighted and the vendor that has been used was the Siemens-TrioTim [19] using a pelvic phased coil. The segmentations were constructed using MeVisLab [20]. The number of image slices for the region of prostate ranged from 15 - 22, depending on the patient and slice thicknesses (either 3mm, 4mm or 5mm). The size of the images was 320x320 pixels and resampling has been applied to respect the networks' input requirements.

### B. Preprocessing

The main preprocessing steps used in our methodology are data augmentation, image resampling and image normalization. As the first step, data augmentation was applied on the original data to cope with overfitting issues that may arise from the use of DL pipelines. Data augmentation was performed using the following affine transformations:(i) rotation of the images in different fixed degrees (-20,-10,-5,5, 10, 20), and (ii) image shifting upwards, downwards, left or right, by a factor of 0.5. Another preprocessing step was image resizing. In particular, after cropping, the images were upsampled using nearest neighbor interpolation. Upsampling is necessary, as a larger number of pixels provides the model with more features to extract information from [21] and also is a technical requirement for the chosen model architectures. Finally, image normalization was performed. DL models need to have their input data normalized to reach a sufficient convergence point [22]. There are plenty of normalization strategies such as the z-score or the minimum-maximum (Min-Max) normalization. We employed the Min-Max normalization slice-wised which applies a linear transformation on the original range of the data. This specific technique fits the data in a pre-defined boundary. The normalized data are computed using the following equation:

$$x' = \left( \frac{x - \min(x)}{\max(x) - \min(x)} \right) \quad (1)$$

where $x'$ is the normalized data distribution and x is the original data distribution and the updated intensity range of the pixels was between 0-1. The data normalization step was applied on the testing data as well in order to achieve the homogeneity which is necessary for the models to provide robust results.

### C. Proposed pipeline for smart cropping

In the present work we propose a method for handling class imbalance through a smart-cropping technique of the central prostate gland and a small region around that area. First, a bounding box enclosing the original mask of the prostate gland was created by expanding the original mask by 40 pixels both vertically and horizontally. Then, the training images were cropped around the area defined by the bounding box. Fig. 1 depicts the original MRI image, the original prostate mask and the bounding box created after the expansion.

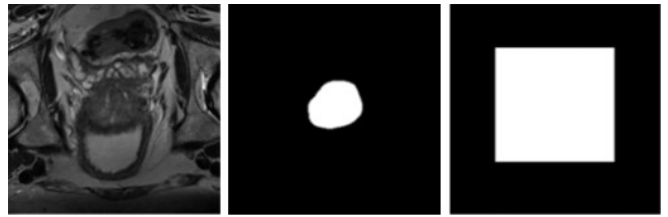

Figure 1: An example of prostate MRI (left), the original mask of the prostate gland (middle) and the bounding box used for cropping (right).

Subsequently, a U-net network [10] was trained to crop those patches from the testing dataset. The U-net architecture is practically an encoder-decoder network for segmentation tasks. Our work targets to tackle class imbalance by cropping unnecessary information around the area of interest without making any a priori assumptions regarding the location of the prostate in the image (i.e. at the image center).

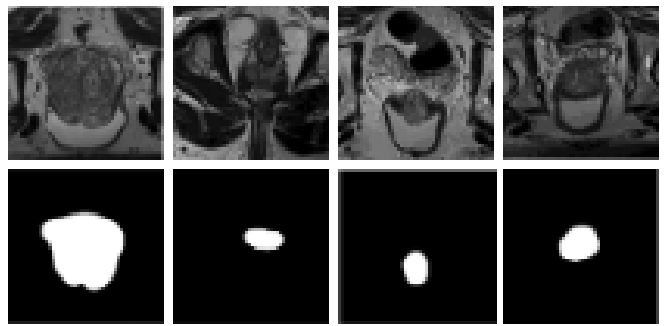

Figure 2: Size variation of the central prostate zone for 4 patients.

As opposed to the typical object localization network where only 4 points at the corners of the bounding box are detected, with the proposed smart-cropping method the network is trained to segment the area around the central zone of the prostate with higher accuracy and a more balanced distribution between foreground and background pixels. The utilization of a neural network for strategical cropping is crucial due to the size variation of the central zone of the prostate gland. An indicative example of the variability in prostate gland size is provided in Fig. 2.

The entire pipeline used to define the bounding box is presented in Fig.3. In steps 1a & 1b a sample from the center-cropped images and the corresponding annotation are shown.

These images are fed into the U-net architecture [10] (step 2), which is trained with the annotation boxes for the prediction of a bigger amorphous area around the mask (step 3). The bounding boxes are then defined (step 4) using the minimum and maximum coordinates on the x and y axis of the amorphous masks (from step 3). This process ensures that the original annotations are always included in the final cropped image. Then, in step 5, resampling is applied to upsample the images into 256x256 pixels. The size of the cropped images before upsampling was approximately 106x117. Finally, in steps 6a & 6b the resulting cropped images along with their cropped annotations are depicted. The resulted images and annotations can be used for training the networks to segment the central prostate gland.

*D. Network training*

The output of the proposed approach (steps 6a & 6b in Fig. 3), was used for training and testing the five segmentation architectures. A brief description of the networks is provided in section E. The architectures were trained and tested using two pipelines. In the first, standard approach, the networks were trained using the images after conventional center-cropping (Fig.3, steps 1a & 1b). For the second pipeline, the proposed smart-cropping approach was applied to crop the images and the resulting images (Fig.3, steps 6a&6b) were used to train the networks.

To ensure that the generalizability of the network is sufficient, training was performed using a 5-fold cross validation strategy [23]. In each fold, 24 patients were used for the training process while the remaining 6 patients were utilized for testing. The patients were partitioned in the folds in the same way for all the architectures and for both smart-crop and center-crop approaches. The overall performance of the trained networks was computed by averaging the 5-fold cross validation results over the five test-sets.

The total number of images (2D slices) in each training fold was approximately 330 before and 630 after data augmentation with image size 256x256 pixels. The number of images used in each testing fold was roughly 90 and no data augmentation was used for the testing dataset.

For the training process, training accuracy and binary cross-entropy loss have been utilized. The Optimization method used is the Adam [24] instead of Stochastic gradient descend because it converges faster. The model was trained for 150 epochs for almost all architectures. Checkpoint strategy and early stopping have been used to reduce the computation time.

*E. Networks for comparison*

In total, five DL architectures have been used to compare our preprocessing approach (smart-cropping) with the typical center-cropping technique: the U-net [10], U-net+, ResU-net++ [25], Bridged U-net[26] and DenseU-net[27]. These architectures have the backbone of U-net architecture which is an encoder-decoder network and it has proved to be the state of the art in segmentation tasks. They were selected for the present study based on algorithm relevance, availability and reproducibility, as the implementations of 4 out of 5 networks (except U-net+) are readily accessible and they have been already applied in the same or closely-relevant application such as ResU-net++ which has been used for polyp segmentation. U-net+ is an original network which is very similar to ResU-net++ without the residual connections. In the following sections we provide a brief description of these networks.

*1) U-net*

The U-net architecture has been proposed by Ronnenberger et al. [10] and the novelty of the work consists of the encoder-decoder path, that can transfer the information from the downsampling path into the upsampling path in a serial and a parallel way, to increase the ability of the network to learn spatial features. Although the typical convolutional networks can analyze the image and find connections between features, they lose location information about the area of interest. Connecting the decoder path with the encoder path the information about the location of the object of interest is passed into the serial path which has the information about the content of the object of interest. In conclusion the parallel paths transfer the "where" information while the serial paths transfer the "what" information. This network encoder path consists of convolutional layers [28] for the feature extraction process max pooling layers [29] for dimensionality reduction which makes the network work efficiently with less computational cost while the encoder path consists of upward convolutional layers [30,31] for the reconstruction of the original dimensions and the reduction of feature channels. Final layer uses a sigmoid function when the output is a binary mask or a softmax function for multi-label segmentation [32–34].

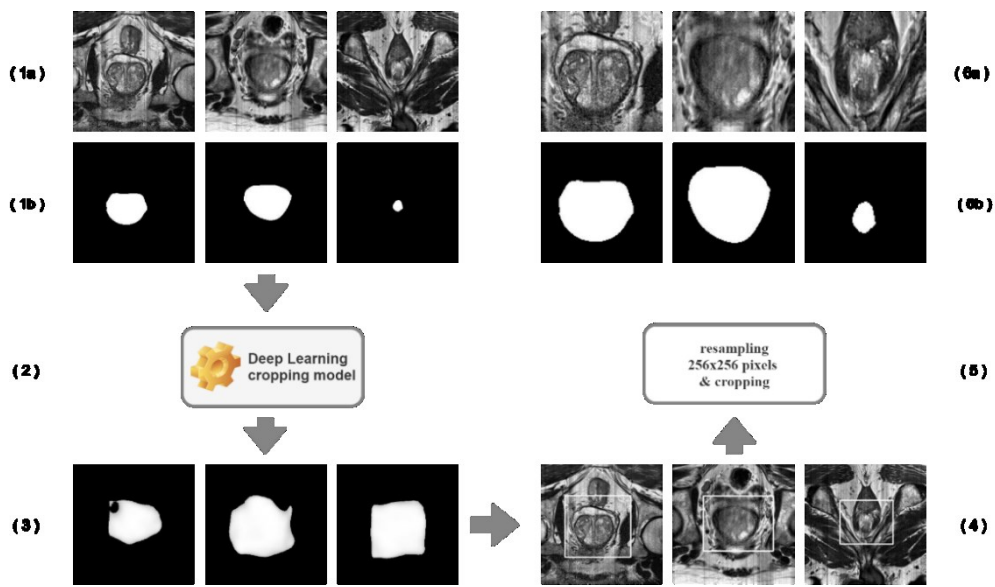

Figure 3: Pipeline for the proposed smart-cropping technique.

*2) ResU-net++*

ResU-net++ is also an encoder-decoder architecture and has already been used for polyp segmentation task [25] where three new layers have been added in the backbone of the U-net. Those blocks are attention networks into the original network's layers [35], squeeze and excitation blocks [36] and the Atrous spatial pyramid pooling layers [37] which have been used in the base and before the output of that network. Briefly, attention networks have been responsible for giving attention to certain significant features of the image and practically enhance the quality and the representation of the features that boost the results making the network smarter in a way. The squeeze and excitation blocks recalibrate the features extracted from the channels, so they are capable of identifying patterns between different feature channels. The Atrous spatial pyramid pooling captures the contextual information in several scales [38] and opens up the receptive field of the network using dilated convolutions [39]. Furthermore, residual connections proposed by He et al. have been used to tackle the degradation problem [40,41].

*3) U-net+*

U-net+ is quite similar to ResU-net++ with the difference that here no residual connections have been used. Attention networks along with Atrous spatial pyramid pooling and squeeze and excitations blocks have been embedded into the networks.

*4) DenseU-net*

DenseU-net [27] is also an encoder-decoder network where, like residual connections, dense blocks [42] are being used to enable the passing of information from previous layers forward while the transitional blocks are reducing the computational burden of the network making the features produced by dense blocks simpler.

*5) Bridged U-net*

Bridged U-net proposed by Chen et al. [26] consists of 2 stacked U-nets and apart from the connections between encoder and decoder paths, there are also inter-network connections between layers from the first U-net to the second. It is important to mention that this architecture has also been used for the segmentation of prostate's central zone.

*F. Evaluation metrics*

To evaluate the segmentation performance on the test sets we used: (a) the Dice score coefficient [43,44] which measures the overlap proportion of the predicted mask over the ground truth mask and (b) the Rand error index [45,46] which is a metric of similarity for data clustering techniques and it has been proposed as a measure of segmentation performance due to the fact that segmentation could be regarded as a clustering of pixels. The mathematical description of the Dice score and the Rand error index, is provided in Eq.(2) and Eq.(3), respectively.

$$Dice\ coeff = \frac{2|y \cap \hat{y}|}{|y| + |\hat{y}|}, \quad (2)$$

$$Rand\ Error\ Index = \frac{y + \hat{y}}{\binom{N}{2}}, \quad (3)$$

where $\hat{y}, y$ are the samples from distributions, $\hat{Y}, Y$ which are the prediction and the ground truth mask, respectively and $N$ is the total number of pixels. Distribution $\hat{Y}$ describes all the predictions from the images while $\hat{y}$ describes a single prediction from an image. Dice score ranges from 0 to 1 with higher values indicating better overlap between segmentation and ground truth. Rand Error index tends to 0 if the segmented image is close to the ground truth and tends to 1 when the difference between both images is important.

*G. Implementation set-up*

The implementation of the former pipelines has been developed in the Python language [47] with the Keras framework [48] for the Deep learning pipelines working on the Tensorflow backend. The versions of the aforementioned packages were 2.3 and 2.2, respectively. The hardware used is an Nvidia Quadro P6000 graphic card, an Intel Core i7-5820k CPU working on 3.3 GHz and RAM of 32 GB. The duration for the whole training process of the proposed pipeline differs between networks but for each of them, it was no more than 10 hours for 5-fold cross validation.

### III. RESULTS

Regarding the class imbalance problem, using the center-cropping approach the ratio of background over foreground pixels was $\frac{background\ pixels}{foreground\ pixels} = 55.53$. The percentage of foreground pixels in the image was on average $\frac{foreground\ pixels}{total\ pixels} = 3.39 \pm 2.05\%$. On the other hand, using the proposed smart-cropping approach, the ratio of background pixels over foreground pixels was reduced to $\frac{background\ pixels}{foreground\ pixels} = 15.38$. The percentage of foreground pixels in the image was on average $\frac{foreground\ pixels}{total\ pixels} = 9.09 \pm 4.34\%$. It is worth mentioning that a completely balanced distribution between foreground and background pixels is not suited for medical applications due to the importance of the background content and its relationship with the objects of interest.

Table I shows the segmentation performance for the five networks using standard center-cropping and the proposed smart-cropping. The mean Dice scores and Rand Error for each architecture for the center crop and the smart crop approaches along with their p-values are given. The Wilcoxon signed-rank test, was used to compare pairwise the two methods. Overall, the proposed pipeline outperformed the center-cropping method for all five architectures. In terms of Dice score, the improvement in segmentation performance using smart-cropping was significant for three out of five networks: U-net, ResU-net++ and U-net+. The highest improvement was achieved with the U-net+ and ResU-net++ architectures corresponding to 8.9% and 8%, respectively. In terms of Dice score, the improvement in segmentation performance using smart-cropping was significant for three out of five networks: ResU-net++, U-net+ and Bridge U-net. It is worth mentioning that Dense U-net outperformed all other networks in segmenting prostate gland both in terms of Dice score (>0.82) and Rand Index (<0.16) regardless of the cropping technique used. This is also the only network for which no significant improvement was achieved using the smart-cropping technique (Table I).

TABLE I
SEGMENTATION PERFORMANCE USING CENTER-CROPPING AND SMART CROPPING FOR 5 ARCHITECTURES

| Metric | Cropping technique | U-net | ResU-net++ | U-net+ | DenseU-net | Bridged U-net |
|---|---|---|---|---|---|---|
| Dice Score | Center cropping | 0.74±0.04 | 0.74±0.05 | 0.71±0.07 | 0.82±0.01 | 0.78±0.03 |
| | Smart cropping | 0.79±0.02 | 0.82±0.02 | 0.79±0.01 | 0.84±0.01 | 0.81±0.02 |
| P-value | | **0.0001** | **<0.0001** | **<0.0001** | 0.54 | 0.85 |
| Rand Error Index | Center cropping | 0.22±0.03 | 0.21±0.03 | 0.24±0.04 | 0.16±0.02 | 0.19±0.02 |
| | Smart cropping | 0.19±0.01 | 0.17±0.01 | 0.20±0.01 | 0.15±0.01 | 0.18±0.01 |
| P-value | | 0.64 | **<0.0001** | **0.021** | 0.89 | **<0.0001** |

Segmentation performance was also computed specifically for the prostate apex (first slice), mid-gland (central slice) and base (last slice) parts of the prostate. The corresponding boxplots of the Dice score and Rand index for the five networks using center- and smart-cropping are shown in Fig. 4 and Fig. 5, respectively. Although the results are comparable for the two cropping methods for prostate midgland, using smart-cropping significantly improves the performance of the segmentation DL networks in the first and last axial slices of the prostate, corresponding to the base and apex parts, respectively.

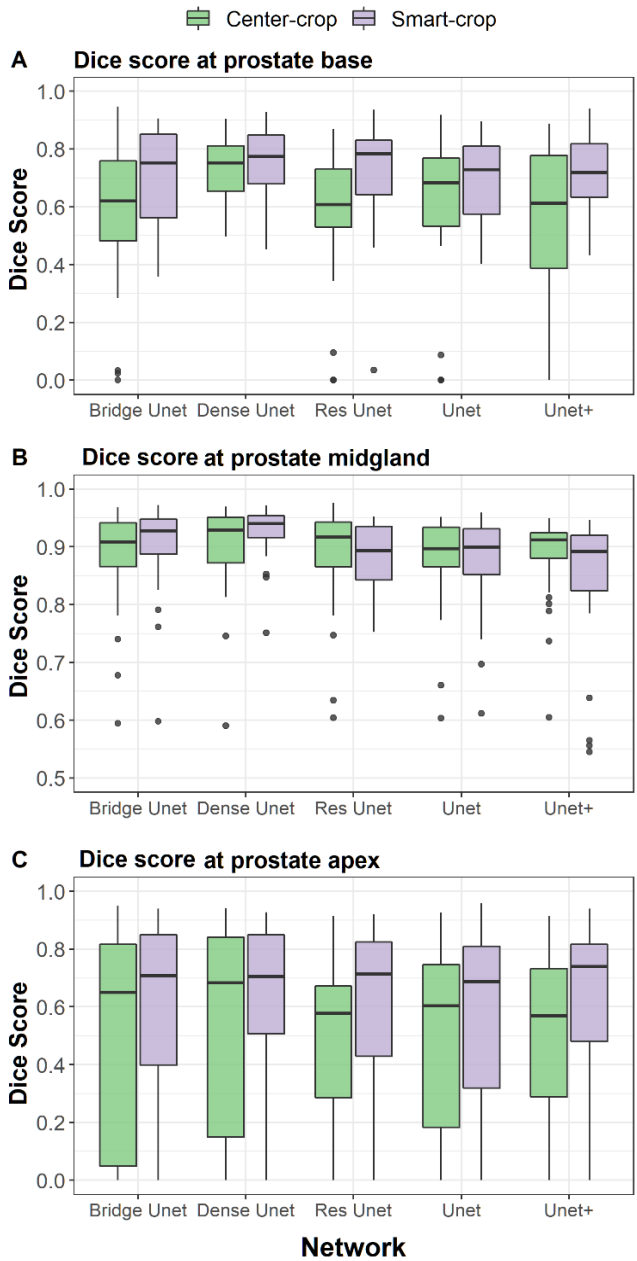

Figure 4: Boxplot of Dice score at prostate base (A), midgland (B) and apex (C) for the five networks using center- and smart-cropping

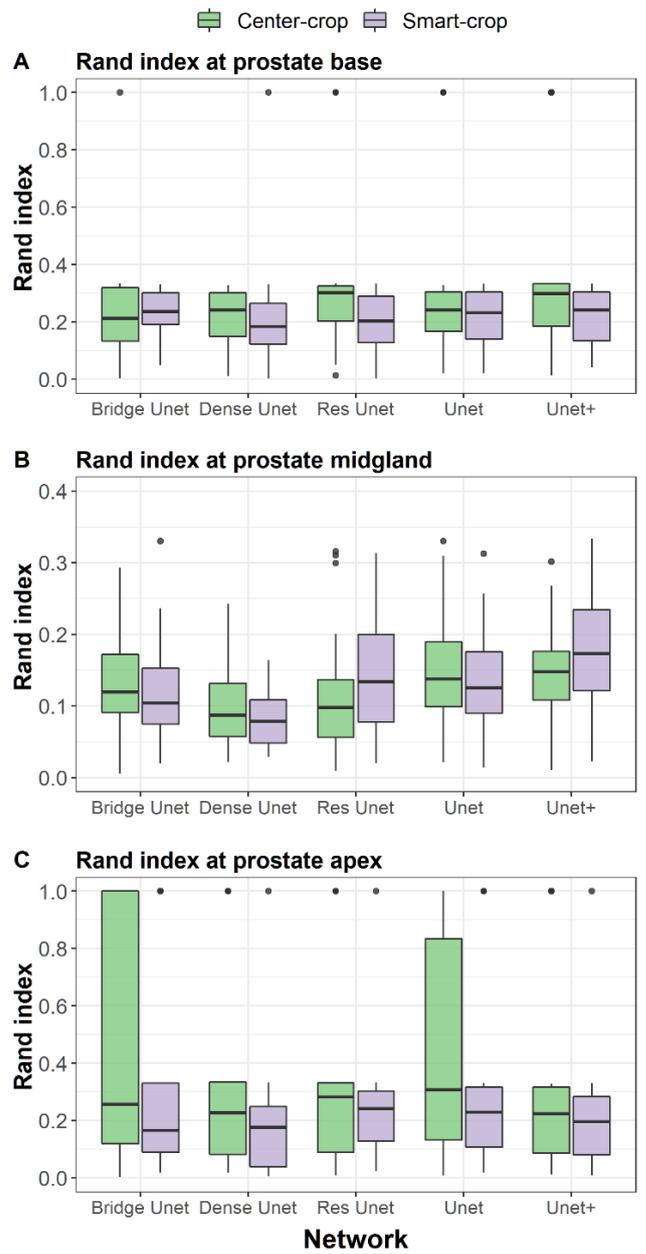

Figure 5: Boxplot of Rand index at prostate base (A), midgland (B) and apex (C) for the five networks using center- and smart-cropping

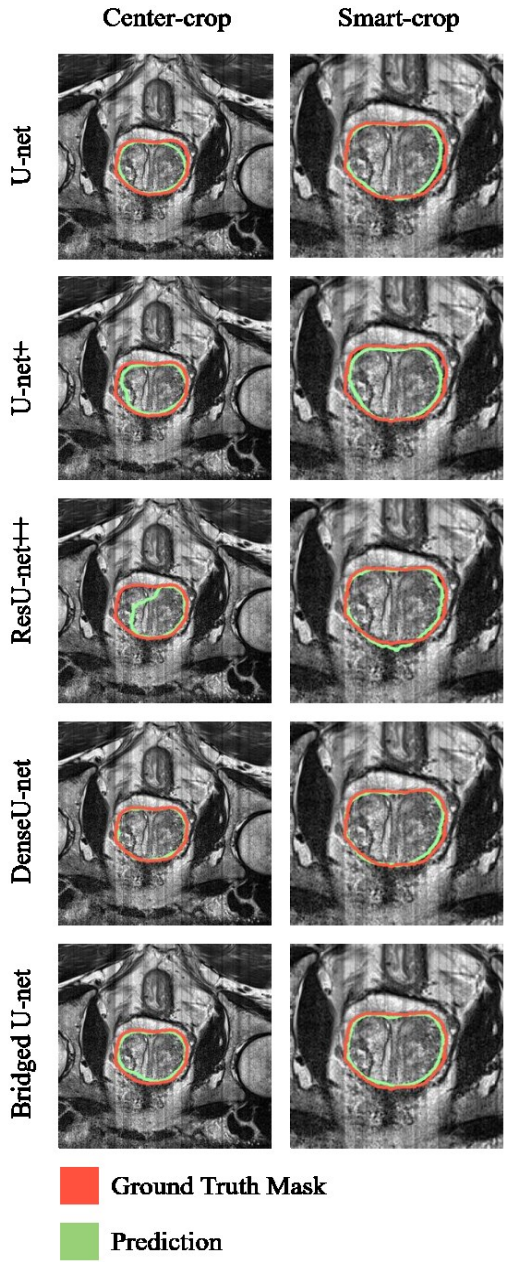

Figure 6: An example of prostate segmentation using center cropping (left colum) and smart cropping (right column) for each one of the five DL architectures

An example of the segmentation result is depicted in Fig. 6. Red contours indicate the original marks of the prostate and green contours indicate the segmented region as predicted with the different architectures for both center- and smart-cropping.

## IV. DISCUSSION AND CONCLUSION

In the present study we propose a novel DL-based architecture for cropping MRI images around the prostate aiming to tackle the class imbalance problem in pixel distribution between foreground (prostate) and background pixels. We demonstrated that the proposed smart cropping technique outperformed the standard center cropping for all the prostate segmentation DL networks used for evaluation both globally, at the entire prostate gland, and locally at the prostate base and prostate apex.

It is a common problem of automatic segmentation methods to over-segment or under-segment the prostate at the upper (apex) and lower (base) parts as these are the most difficult parts to segment due the large variability and slice thickness. Indeed, every algorithm performed worse on the apex and base compared to the mid-gland if we look at the metric values in Fig.4 and Fig.5. Nevertheless, for several applications, for example in radiotherapy and TRUS/MR fusion, it is critical to segment correctly the apex and the base of the prostate [49]. Therefore, it is of paramount importance to evaluate segmentation performance not only at the entire prostate gland but also locally at the regions prone to errors. In the present study, in addition to evaluating segmentation performance over the entire set of 2D axial slices of the prostate, we also calculated segmentation performance specifically for the apex, mid-gland, and base of the prostate. Interestingly, the proposed smart-cropping method appears to effectively reduce the apex and base segmentation ambiguities compared to standard center-cropping.

Class-imbalance in machine learning has been widely studied, however, little attention has been paid on the subject of object detection. Particularly in image segmentation tasks, the issue arises mainly from the usage of cross entropy (CE) as a loss function, as it is well known in literature that CE has difficulty in handling class imbalance [50]. The CE loss is applied on each image separately during the training process and the weights of the network are tuned based on the mean results of CE after a batch of images has passed from forward propagation. This can be problematic if the classes have unbalanced representation in the image, as the most prevalent class can dominate training [51]. To alleviate this problem, the basic approach is to assign weights to classes based on the inverse of their occurrences but the choice of weighting is non-trivial and application-dependent. Commonly, the weighted cross-entropy loss is used to counteract a class imbalance present in an imaging dataset [52]. This approach has two drawbacks: 1) assigning a proper weight will be an issue for a dataset with varying object sizes, 2) the least frequent class will be affected by noise and it may result in unstable training.

To tackle the class imbalance problems, several types of loss function have been utilized individually or combined in medical image segmentation tasks. Although the weighted CE loss function has been utilized for tackling the class imbalance issue, there are still several limitations regarding this method. First, by adding weights to a certain class during training, the model is adjusted to the data which automatically adds bias into the model. Second, fitting the model to the data using weighted CE makes the model more sensitive to the distribution of the data which means that if the testing set's distribution is slightly different from that of the training set, then the model will be ρnegatively affected. Lin et al. proposed a novel focal loss [53] which is defined by introducing a modulating factor to the CE loss to differentiate between background and foreground pixels. While focal loss has shown promising results compared to CE, it faces difficulty with datasets having severe class imbalance. Another study has demonstrated that metric-sensitive losses are superior to CE-based loss functions when the Dice Score or Jaccard Index are used for evaluating segmentation performance [54]. Furthermore, metric-sensitive losses are invariant to scale and, therefore, are expected to perform better with segmentation of small-sized objects [55]. Further research is needed to assess the improvement of smart cropping in segmentation performance with respect to state-of-the-art loss functions.

The major limitation of our study is the number of patients available for the study which was not large enough to reproduce the results provided in the literature. For instance, state-of-the-art results for DenseU-net architecture [27] demonstrate a Dice score of 90% for the central prostate zone while with the center crop approach used in the present study (which is similar to the one used in [27]) the Dice score was 84%. More data are also needed to improve the robustness of the proposed approach.

To summarize, the proposed optimized DL-based smart copping technique that improves class balance around the prostate may significantly improve prostate gland segmentation compared to standard center cropping. To establish the generalizability of our method, these findings need to be validated in external independent populations including images taken from different MRI vendors.